\documentclass{aa}

\usepackage{longtable,lscape}
\usepackage{graphicx}
\usepackage{natbib}
\usepackage[modulo,switch]{lineno}
\bibpunct{(}{)}{;}{a}{}{,}
\usepackage{txfonts}
\usepackage{color}

\begin{document}
\def\cd{d$^{-1}$}
\def\cds{d$^{-1}$\,}
\def\kms{km\,s$^{-1}$}
\def\fu{$f_1$}
\def\fd{$f_2$}
\def\ft{$f_3$}
\def\fq{$f_4$}
\def\fc{$f_5$}
\def\fB{$f_m$}
\def\hads{CoRoT~101155310}
\def\Hads{High--amplitude $\delta$ Sct}
\def\HADS{high--amplitude $\delta$ Sct}

    \title{Monitoring a high--amplitude $\delta$ Sct star 
for 152~days:
discovery of 12 additional modes and modulation effects in the light curve of \hads
\thanks{
The CoRoT space mission was developed and  is operated by the French
space agency CNES, with the participation of ESA's RSSD and Science Programmes,
Austria, Belgium, Brazil, Germany, and Spain.
This work uses ground--based spectroscopic observations made with the HARPS instrument
at the 3.6m-ESO telescope (La Silla, Chile) under the ESO Large Programme LP182.D-0356 and
complementary photometric measurements made at
the Piszk\'estet\H{o} Mountain Station of Konkoly Observatory (Hungary).}$^{,}$
\thanks{Tables 1 and 2 are available in electronic form
at the CDS via anonymous ftp to cdsarc.u-strasbg.fr (130.79.128.5)
or via http://cdsweb.u-strasbg.fr/cgi-bin/qcat?J/A+A/vol/page}}
   \author{
E.~Poretti\inst{1} 
\and
M.~Rainer\inst{1}
\and 
 W.W.~Weiss \inst{2}
\and
Zs.~Bogn\'ar\inst{3}
\and
A.~Moya\inst{4}
\and
E.~Niemczura\inst{5}
\and
J.C.~Su\'arez\inst{6}
\and
M.~Auvergne\inst{7}
\and
A.~Baglin\inst{7}
\and
F.~Baudin\inst{8}
\and
J.M.~Benk\H{o}\inst{3}
\and
{J.~Debosscher}\inst{9}
\and
R.~Garrido\inst{6}
\and
L.~Mantegazza\inst{1}
\and
M.~Papar\'o\inst{3}
 }
   \authorrunning{Poretti et al.}
\titlerunning{Nonradial modes and modulation effects in \hads}

   \institute{ INAF -- Osservatorio Astronomico di Brera,
              Via E. Bianchi 46, 23807 Merate (LC), Italy\\
              \email{ennio.poretti@brera.inaf.it} 
\and Institute of Astronomy, University of Vienna, T\"urkenschanzstrasse 17, A-1180 Vienna, Austria
\and Konkoly Observatory, PO Box 67, 1525 Budapest, Hungary
\and Departamento de Astrof\'{\i}sica, CAB (INTA-CSIC), PO Box 78, 28691 Villanueva de la 
     Ca\~{n}ada, Madrid, Spain
\and Instituto de Astrof\'{\i}sica de Andaluc\'{\i}a, Apartado 3004, 18080 Granada, Spain
\and Astronomical Institute of the Wroclaw University, ul. Kopenika 11, 51-622 Wroclaw, Poland
\and LESIA, Universit\'e Pierre et Marie Curie, 
    Universit\'e Denis Diderot,  Observatoire de Paris,  92195 Meudon, France
\and Institut d'Astrophysique Spatiale, CNRS, Universit\'e Paris XI UMR 8617, 91405 Orsay, France
\and Instituut Voor Sterrenkunde, Catholic University of Leuven, Celestijnenlaan 200D, 3001
     Leuven, Belgium
          }
   \date{Received, accepted}
   \abstract
   {}
{The detection of small--amplitude nonradial modes in high-amplitude $\delta$ Sct (HADS) 
variables has been very elusive until at least five of them were detected in the
light curve of V974 Oph obtained from ground--based observations.  The combination of radial and nonradial modes has
a high asteroseismic potential, thanks to the strong constraints we can put in the
modelling. 
The continuous monitoring of ASAS 192647-0030.0$\equiv$\hads\, ($P$=0.1258~d, $V$=13.4)
ensured from space by the CoRoT (Convection, Rotation and planetary Transits) 
mission constitutes a unique opportunity to exploit such potential.}
{The 22270 CoRoT measurements were performed in the chromatic mode. They span
152~d and cover 1208 consecutive cycles.
After the correction for one jump and the long-term drift,
the level of the noise turned out to be  29~$\mu$mag. 
The phase shifts and amplitude ratios of the coloured
CoRoT data, the HARPS spectra, and the period-luminosity relation  were used to determine
a self--consistent physical model. In turn, it allowed us to model the oscillation
spectrum, also giving  feedback on the internal structure of the star.
 }
   {In addition to the fundamental radial mode \fu=7.949~\cds with harmonics up to 10\fu, we detected 12 
independent terms. Linear combinations were
also found and  the light curve was solved by means of 61 frequencies (smallest amplitude
0.10~mmag). The newest result is the detection of a periodic modulation
of the \fu\, mode (triplets at $\pm$0.193~\cds centred on \fu\, and 2\fu), discussed as
a rotational effect or as an extension of the Blazhko effect to HADS stars.
The physical model suggests that \hads\, is an evolved star, with a slight subsolar
metallic abundance, close to the terminal age main
sequence. All the 12 additional terms are identified with mixed modes in the 
predicted overstable region.
 }
{}
    \maketitle
%

\section{Introduction}
\Hads\, (HADS) stars occupy the central part of the lower part of the instability strip between
the horizontal branch and the main sequence \citep{vienna}. In the same region of the
instability strip, 
low--amplitude $\delta$ Sct stars span a wider range of temperatures and luminosities.
The phenomenological separation is based on an amplitude limit of 0.20-0.30 mag, but 
actually it has never been considered as  physically significant. Moreover, 
HADS stars could not be further considered  mono-- or double--mode  radial pulsators after 
detecting  additional modes \citep[][and references therein]{oph}. Their 
identification as nonradial modes opened the possibility of considering HADS stars
as new tools for asteroseismic studies.
Indeed, recent theoretical investigations \citep{pet2,pet1}
indicate that rotation effects may provide
useful additional information on the mode identification, in particular
for the nonradial modes rotationally coupled with the fundamental
and first-overtone radial modes.
That large amplitude modes could be  identified radial allowed us to use
HADS stars  as distance indicators \citep{fornax,macho}.
The most luminous HADS stars are just below the
horizontal branch, and  the debate whether the light curves of HADS stars could show
similarities with those of  RR Lyr stars is still open \citep{blabre}.

A great improvement in the observational techniques and data quality was needed
to investigate both the pulsational content and the behaviour of the predominant radial
modes. This opportunity could only be offered by a space mission.

\section{The CoRoT data}
The variability of the 13$^{\rm th}$--mag star located at 
$\alpha$ = $19^{\mathrm h}$\,$26^{\mathrm m}$\,$47\fs{25}$,
$\delta$ = --00\degr\,30\arcmin\,01\farcs{44} (J2000)
was discovered during the ASAS survey \citep[All Sky Automated Survey;][]{asas}.
The star is named 192647-0030.0 and classified as a $\delta$ Sct variable
with $P$=0.1258~d in the ASAS catalogue ({\tt http://www.astrouw.edu.pl/asas/}).
This part of sky, i.e., the equatorial plane not far from the galactic centre, 
is well within one of the ``eyes" of the CoRoT space mission
\citep[COnvection, ROtation and planetary Transits;][]{esa3,flight}.
The inclusion of ASAS~192647-0030.0  in the exoplanetary field of the first
 long run in the direction of the galactic centre (LRc01) was done after searching the catalogue
compiled by \citet{eyes}.
Therefore, we specifically requested the CoRoT monitoring
of  ASAS~192647-0030.0 as part of the 
 Additional Programmes \citep{esa4}.
In the CoRoT database the star was named \hads$\equiv$USNOA2 0825-15821309 
\citep[$V$=13.43, $B-V$=+0.60, $V-R$=0.21;][]{exodat}.
There is no relevant contamination from nearby stars, since the brightest
star included in the CoRoT mask is 2.9~mag fainter than \hads\, in $V$ light
\citep{exodat}.  
The ``CoRoT Variability Classifier" automated supervised method \citep{cvc}
classified it as an RR Lyr star only on the basis of the Fourier parameters.
Since some  overlap is present with the HADS stars in this parameter
space \citep{ogle}, the classification system has been consequently  
revised to better separate RR Lyr from HADS stars.

The LRc01 started on May 15, 2007 and finished 152~d later, on 
October 14, 2007. This means that \hads\, was monitored for
1208 consecutive cycles. The exposure time in the CoRoT exoplanetary
channel was 512~s. This value remained constant over the long run, since
no regular feature suggesting planetary transit was noticed. The Heliocentric
Julian Dates (HJD) were reported to the times at half exposure.
The very high duty cycle allowed us to collect 25597 datapoints.
\hads\, was observed in the chromatic mode,  by means of
 a bi--prism located in the focal block of the instrument, between
the dioptric objective and the plane of the detectors \citep{esa5}. The image is a small, 
low--resolution spectrum whose area is about 300~arcsec$^2$ on the sky.
The photometry is performed in three different colours, roughly red ($r_C$),
 green ($g_C$), and blue ($b_C$), to supply a tool able to  
distinguish between planetary  transits 
(colour independent) and stellar activity (changing with the wavelength).
The white  light, i.e., the sum of the three contributions, is more suitable for 
signal detection thanks to its  larger number of counts.  
\hads\, was also observed with the 1.0~m telescope of  the Konkoly Observatory at the 
the Piszk\'estet\H{o} Mountain Station (Hungary) on two nights in June 2009. 
CCD  photometry with standard $BVI$ filters was planned to 
compare the amplitudes in the CoRoT passbands with those in standard ones. 
The 88 measurements in each filter cover one complete  pulsational cycle, 
and we could also determine one time of maximum brightness.

\begin{figure}[]
\begin{center}
\includegraphics[width=\columnwidth,height=\columnwidth]{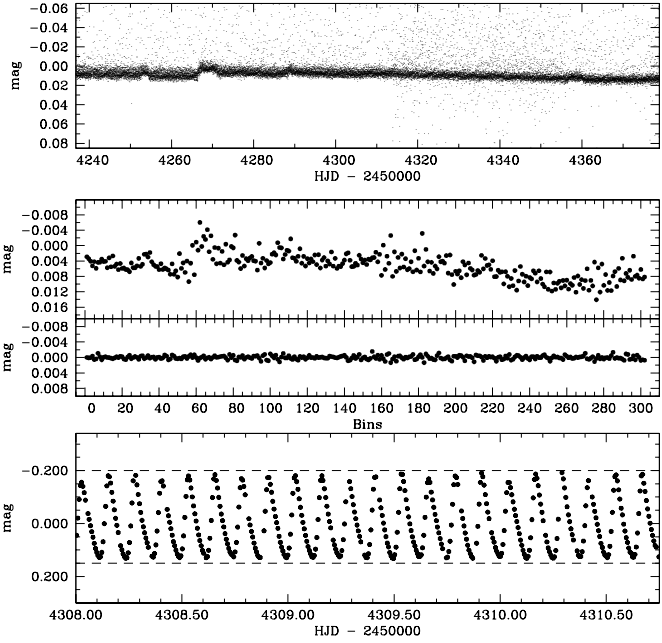}
\caption{\footnotesize
{\it Top panel:} curve of the residuals (a preliminary solution was subtracted from the
original data) showing the long-- and short--term
instrumental effects. {\it Middle panel:} behaviour of the mean magnitudes before (upper curve)
and after (lower curve) the removal of the instrumental effects shown in the upper panel. Each
bin spans four pulsational cycles. {\it Bottom panel:} the example extracted from the final timeseries
of \hads\,  shows the continuous monitoring obtained from space by CoRoT.}
\label{drift}
\end{center}
\end{figure}

The CoRoT measurements are a sort of absolute photometry, since the measured  
flux is  not corrected with respect to any reference source or star. Therefore,
instrumental changes and drifts have a direct influence on  it, on both  
short- and long-term scales. 
Different detrending algorithms can
be used, based on  moving means or polynomial fits. After several trials, we were
more satisfied by an approach based on a preliminary frequency analysis. Indeed, the
main frequency $f_1$=7.949~d$^{-1}$ and other high--amplitude terms were immediately 
detected in the raw data. 
We calculated the least--squares fit of the raw data
by using a preliminary solution based on the terms with the largest amplitudes.
The residuals in  white light (Fig.~\ref{drift}, upper panel) show a long--term drift
and more rapid and apparently irregular changes around JD 2454253 and  2454267. Moreover, in the 
second half of the observations the distribution of the outliers strongly changed and
many points appeared well below the others. This so-called ``rain" of points from the middle strip
has an instrumental origin 
because it is observed in other stars belonging to the same pointing. 
To correct the irregular behaviour of the mean magnitude
we calculated  the  least--squares fits on four consecutive cycles.
These fits returned  the 302 values
of the mean magnitudes (Fig.~\ref{drift}, upper curve in the middle panel), which mimics
the trend of the residuals. 
The standard deviations of the 302 solutions range from 0.004 to 0.016 mag.
To remove the instrumental
variations, we interpolated the values of the mean magnitudes at the time
of each individual measurement of \hads\, and then subtracted it.

After that, we recalculated the least--squares fits on the
302 bins and  obtained very homogeneous values of the mean magnitudes (Fig.~\ref{drift}, 
lower curve in the middle panel). 
By applying this procedure  we also removed any oscillation around 0.5~d, 
but as we see later (Sect.~\ref{duecd}) this did not have a strong 
impact on our solution.
At this stage we also removed the outliers.
A post--facto analysis of the raw and the
cleaned datasets showed that the frequency content of the two timeseries is 
the same.
The bottom panel of Fig.~\ref{drift} shows an example of the excellent quality of the 
final timeseries and also allows us to appeciate the continuous CoRoT monitoring.

The same technique was applied to the coloured data.
The trend in $r_C$ light is roughly the same as in white light (top panel of Fig.~\ref{drift}),
since the total flux in this light (about 182000 adu) is much greater than of $g_C$ 
(about 22600 adu) and $b_C$ (about 32300 adu). The main difference is in the
decline   after JD 2454310, which is sharper in $r_C$ light  than
in white light. Actually, the fluxes observed in the $g_C$  and $b_C$ light
increased after this date, and what we measured in white light is the total flux variation.
This changeover concerns those days when the ``rain" effect appeared, and it is very probably
a jitter effect through the photometric mask of \hads.

\section{The frequency content}
\begin{figure}[]
\begin{center}
\includegraphics[width=\columnwidth,height=\columnwidth]{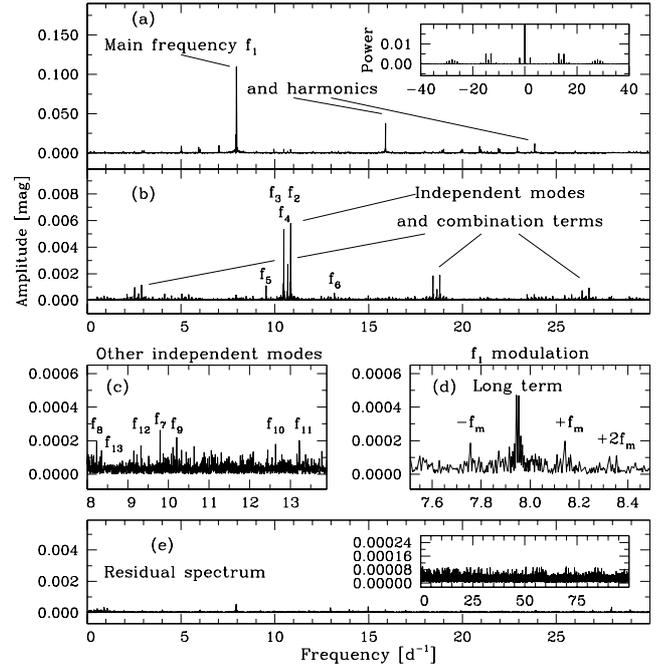}
\caption{\footnotesize Subsequent steps in the detection of frequencies
in the amplitude spectra of \hads.}
\label{ampli}
\end{center}
\end{figure}

The frequency analyses were performed  on different versions of the CoRoT timeseries 
(raw data, detrended data, coloured data, with and without outliers, etc.) and they supplied very
similar results. The main change is in the level of the noise and consequently in the
threshold for the significance of the peaks. Therefore, for the sake of clarity, we discuss
here the detection performed in white light.
The frequency values were firstly obtained by means of 
the iterative sine--wave, least--squares fitting method \citep{vani} and then
refined by the MTRAP algorithm \citep{mtrap}, both at intermediate stages and at the end of the
whole analysis. 

The level of the noise in the final timeseries (22270 datapoints in white light) 
is 29\,$\mu$mag in the 0-100~\cds region. 
The long--term drift left a slightly higher noise in the 0-5~\cds region, namely 41 $\mu$mag. 
We accepted  61  peaks as having a stellar origin, i.e., not counting  the orbital frequency of the 
satellite $f_{orb}$=13.97~d$^{-1}$, the sidereal frequency $f_{sid}$=1.0027~d$^{-1}$,
 and their harmonics and combinations.  To do that, we firstly
considered the peaks with an amplitude greater than 0.12~mmag, i.e., with a signal--to--noise ratio
(SNR) above 4.0.  Then, we noted that several combination terms (harmonics and coupling
terms) were also detected below this limit. Therefore,  we relaxed the minimum amplitude 
to  0.10~mmag, i.e., 3.5 times the noise level. Some combination terms (e.g., \fu+\fd+\ft, 4\fu-\fq,
3\fu+\fc) are just below the acceptance threshold. 

Figure~\ref{ampli} shows the steps in the solution of 
the CoRoT light curve  of \hads, which led to the detection of
four categories of frequencies:
\begin{enumerate}
\item the predominant term \fu\, and its harmonics; 
\item the independent terms, from  \fd\, to $f_{13}$;
\item the modulation term \fB\, of the term \fu;
\item the combination terms.
\end{enumerate}
\begin{table*}
\caption{ Identification, frequencies, Fourier amplitudes, and phases (T$_0$=HJD 2454236.7502)
of  the terms  detected in the data of \hads\, (22270 datapoints).}
\begin{tabular}{crrr c crrr r}
\hline
\hline
\noalign{\smallskip}
\multicolumn{1}{c}{ID} &
\multicolumn{1}{c}{Frequency} &
\multicolumn{1}{c}{Amplitude} &
\multicolumn{1}{c}{Phase} & &
\multicolumn{1}{c}{ID} &
\multicolumn{1}{c}{Frequency} &
\multicolumn{1}{c}{Amplitude} &
\multicolumn{1}{c}{Phase} & \multicolumn{1}{c}{\fu/$f_i$}
 \\
& \multicolumn{1}{c}{$\mathrm{[d^{-1}]}$} &
\multicolumn{1}{c}{[mag]} &
\multicolumn{1}{c}{[0,2$\pi$]} & &
 &\multicolumn{1}{c}{$\mathrm{[d^{-1}]}$} &
\multicolumn{1}{c}{[mag]} &
\multicolumn{1}{c}{[0,2$\pi$]} & \\
\\
\noalign{\smallskip}
\hline
\noalign{\smallskip}
\multicolumn{4}{c}{Main frequency and harmonics} & &
\multicolumn{5}{c}{Additional terms identified as independent modes}\\
\noalign{\smallskip}
$f_1  $  & \multicolumn{1}{l}{~~7.949172(1)}  &   \multicolumn{1}{l}{0.137632(26)}  &  \multicolumn{1}{l}{5.3775(2)}  &  &
$f_{2}$ & 10.847014(015)  &   0.006224(26)  &  6.0785(0041)  &   0.733         \\
$2f_1 $ & \multicolumn{1}{l}{15.898343}           &   \multicolumn{1}{l}{0.041859}  &  \multicolumn{1}{l}{2.3933}  &  &
$f_{3}$ &  10.479540(016)  &   0.005689(26)  &  5.8511(0045)  & 0.759           \\
$3f_1 $  &  \multicolumn{1}{l}{23.847515}         &   \multicolumn{1}{l}{0.014404}  &  \multicolumn{1}{l}{5.6197}  &   &
$f_{4}$ & 10.690302(030)  &   0.003076(26)  &  5.8388(0083)  &    0.744     \\
$4f_1 $   &  \multicolumn{1}{l}{31.796687}         &   \multicolumn{1}{l}{0.004433}  &  \multicolumn{1}{l}{2.4116}  &   &
$f_{5}$ &  9.532648(075)  &   0.001230(26)  &  3.1547(0208)  &   0.834         \\
$5f_1 $    &   \multicolumn{1}{l}{39.745859}       &   \multicolumn{1}{l}{0.002146}  &  \multicolumn{1}{l}{5.4146}  &   &
$f_{6}$ & 13.175878(147)  &   0.000629(26)  &  1.1226(0407)  & 0.603          \\
$6f_1 $     &  \multicolumn{1}{l}{47.695030}       &   \multicolumn{1}{l}{0.001150}  &  \multicolumn{1}{l}{2.2493}  &   &
$f_{7}$ &  9.796144(340)  &   0.000273(26)  &  5.7221(0937)  & 0.811           \\
$7f_1 $    &  \multicolumn{1}{l}{55.644203}        &   \multicolumn{1}{l}{0.000552}  &  \multicolumn{1}{l}{5.3327}  &   &
$f_{9}$ & 10.203291(385)  &   0.000241(26)  &  4.1313(1062)  &  0.779          \\
$8f_1 $     &  \multicolumn{1}{l}{63.593373}       &   \multicolumn{1}{l}{0.000317}  &  \multicolumn{1}{l}{2.3551}  &   &
$f_{8}$ &  8.232572(450)  &   0.000206(26)  &  0.2385(1242)  &  0.966          \\
$9f_1 $      &  \multicolumn{1}{l}{71.542544}      &   \multicolumn{1}{l}{0.000130}  &  \multicolumn{1}{l}{5.3127}  &   &
$f_{10}$ & 12.635143(455)  &   0.000204(26)  &  0.0702(1254)  & 0.629        \\
$10f_1$       &  \multicolumn{1}{l}{79.491717}     &   \multicolumn{1}{l}{0.000100}  &  \multicolumn{1}{l}{2.6673}  &   &
$f_{11}$ & 13.224629(504)  &   0.000184(26)  &  3.3973(1391)  &   0.601        \\
& & & & &  $f_{12}$&   9.325259(549)  &   0.000169(26)  &  5.2843(1514)  &    0.852      \\
& & & & & $f_{13}$ &  8.354265(672)  &   0.000138(26)  &  1.0701(1854)  &  0.952         \\
\noalign{\smallskip}
\multicolumn{4}{c}{Modulation terms with \fB=0.1933~\cd} & &
\multicolumn{4}{c}{Linear combinations between \fu\, and \fd}\\
\noalign{\smallskip}
$f_1-f_m$ & 7.755895           &   0.000194  &  0.1805  &   &
$f_1+f_2$    & 18.796186       &   0.002261  &  2.2225  &         \\
$f_1+f_m$ & 8.142448           &   0.000145  &  1.9766  &   &
$f_2-f_1$    & 2.897843        &   0.001524  &  3.6718  &        \\
$2f_1-f_m$& 15.705067            &   0.000131  &  2.5886  &  &
$2f_1+f_2$   & 26.745358        &   0.001323  &  5.5501  &      \\
$2f_1+f_m$&    16.091183         &   0.000106  &  4.5329  &   &
$2f_1-f_2$   & 5.051329        &   0.000814  &  5.5587  &       \\
 &             &    &    &   &   $3f_1+f_2$   & 34.694529        &   0.000687  &  2.4496  &       \\
\multicolumn{4}{c}{Linear combinations between \fu\, and \ft} & &
$3f_1-f_2$   & 13.000501        &   0.000219  &  2.8031  &       \\
 &             &    &    &   & $4f_1+f_2$   &  42.643701       &   0.000344  &  5.5241  &      \\
$f_1+f_3 $  & 18.428712         &   0.002247  &  2.1349  &  &$5f_1+f_2$   & 50.592873        &   0.000213  &  2.6688  &      \\
$f_3-f_1 $  & 2.530368         &   0.001398  &  3.4160  &  &$5f_1-f_2$   & 28.898844        &   0.000131  &  4.2861  &       \\
$2f1+f_3 $  & 26.377884        &   0.001104  &  5.3307  &   & $6f_1+f_2$   &58.542046         &   0.000144  &  5.5907  &       \\
$2f1-f_3 $  & 5.418803         &   0.000781  &  5.8036  &  &   \\
$3f1+f_3 $  & 34.327056         &   0.000615  &  2.2546  &  & \multicolumn{4}{c}{Linear combinations between \fu\, and \fc}    \\
$3f1-f_3 $ &  13.367975           &   0.000169  &  2.7857 &      \\
$4f1+f_3 $  & 42.276226          &   0.000303  &  5.3800   & &  $f_1+f_5$   &  17.481819       &   0.000322  &  5.6402  &          \\
$5f1+f_3 $  & 50.225399         &   0.000199  &  2.2301   & & $f_5-f_1$   & 1.583476        &   0.000145  &  0.5953  &        \\
 &             &    &    &   & $2f_1+f_5$  &  25.430991        &   0.000139  &  2.4076  &       \\
\\
\multicolumn{4}{c}{Linear combinations between \fu\, and \fq}& &
\multicolumn{4}{c}{Linear combinations between \fu\, and $f_6$}\\
\\
$2f_4$      &  21.380604    &   0.000164  &  2.8776  &  & $f_1+f_6$   &  21.125050        &   0.000217  &  3.6347  &        \\
$f1+f_4$    &  18.639474    &   0.001137  &  2.0182  &  &  $f_6-f_1$    & 5.226706        &   0.000141  &  4.2040  &     \\
$f_4-f_1$   &   2.741130    &   0.000813  &  3.4594  &  &  $2f_1+f_6$    & 29.074221        &   0.000153  &  0.1586  &     \\
$2f_1+f_4$  &  26.588646    &   0.000620  &  5.3297  &  &  $2f_1-f_6$     & 2.722466       &   0.000135  &  5.2315  &    \\
$2f_1-f_4$  &   5.208042   &   0.000479  &  5.6289  &  &     \\
$3f_1+f_4$  &  34.537817    &   0.000309  &  2.1560  &  &  \multicolumn{4}{c}{Linear combinations between \fd\, and \ft}\\
$3f_1-f_4$  &  13.157213    &   0.000127  &  2.6059  &  &    \\
$4f_1+f_4$  &  42.486989    &   0.000121  &  5.4129  &  &   $f_2+f_3$       & 21.326554     &   0.000194  &  3.3195  &       \\
$5f_1+f_4$  &  50.436162    &   0.000105  &  2.3231  &  &    \\
\noalign{\smallskip}
\hline
\label{sol}
\end{tabular}
\end{table*}
The parameters of the least--squares solution (cosine series) based on the 61 frequencies 
thus detected are listed in Table~\ref{sol}. The residual rms is 0.0027~mag.
The values of the harmonics and combination terms 
were obtained from the  values of the independent frequencies
and from the modulation frequency \fB=0.1933~\cd, for which 
formal errorbars \citep{dsn}  are given between brackets.
This ``locked" solution was calculated by means of the MTRAP algorithm \citep{mtrap}.
The residual spectrum (Fig.~\ref{ampli}, panel {\it e}) shows some peaks 
at the values of the linear combinations of $f_{orb}$=13.97~d$^{-1}$ and
$f_{sid}$=1.0027~d$^{-1}$. These spurious peaks  are due to
the passage of  the satellite over the South Atlantic Anomaly,
which occurred twice daily. 
The environmental conditions  \citep[e.g., the effects of the eclipses on the electronics units,
the eclipse durations, the difference in the Earth's albedo of the overflown regions;][]{flight}
are affecting the CoRoT photometry in such a complicated 
way that, as  in the case of CoRoT 101128793 \citep{793}, the usual
technique of prewhitening  was not able to clean the data in a satisfactory way.

\subsection{The $f_1$ term and its harmonics} \label{ephem}
\begin{figure}[]
\begin{center}
\includegraphics[width=\columnwidth,height=\columnwidth]{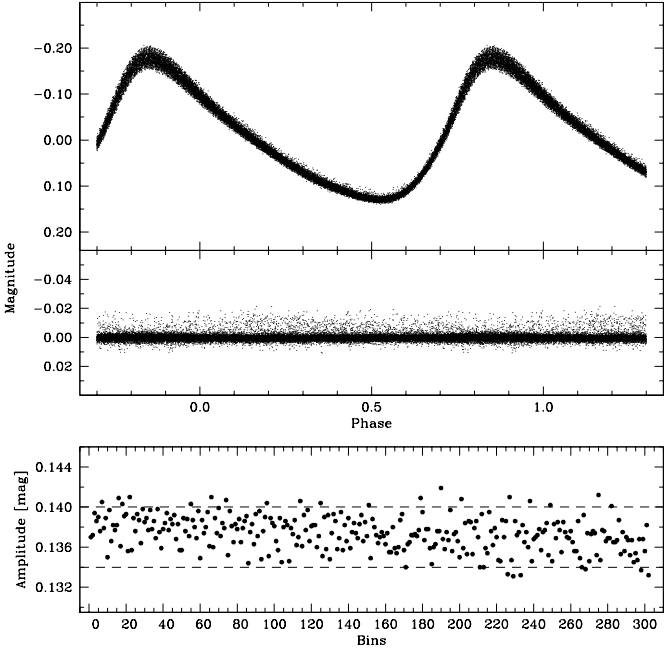}
\caption{\footnotesize
{\it Top panel:} the CoRoT data folded with the main pulsational frequency \fu\,
(upper curve) and the residuals after prewhitening of the 61 frequencies (lower
curve). {\it Bottom panel:} amplitudes of the \fu\, term calculated from measurements
spanning four consecutive pulsational cycles.
}
\label{f1}
\end{center}
\end{figure}

The peaks at \fu=7.949~\cd, 2\fu, and 3\fu\, clearly stand out in the amplitude spectrum
of the original data (Fig.~\ref{ampli}, panel {\it a}).
The spectral window is free of any relevant alias structure, the highest peak
being at the power level of 0.005 (see the inserted box).  
We detected harmonics up to 10\fu\, in the CoRoT light curve, i.e., about twice
those detected in ground--based timeseries. Since the amplitude of harmonics
is monotonically decreasing, the light curve is smoothed (Fig.~\ref{f1}, upper panel), 
without the bumps  observed in RR Lyr stars, both monoperiodic \citep{ogle} and 
multiperiodic \citep{aql,793}. 

The Fourier parameters ($\phi_{21}=4.19$ rad, $\phi_{31}=2.04$ rad, $\phi_{41}=-0.27$ rad, and
$R_{21}=0.38$) are in excellent agreement with those calculated from the OGLE light curves of HADS
stars \citep{ogle}. These parameters suggest that \hads\, is pulsating in the
fundamental radial mode.

We calculated the times of the maximum brightness  by fitting the measurements
of the well--covered cycles with a cubic spline.
In such a way we could determine 1090 times of maxima 
(Table~2) and in turn  the  ephemeris
$
\begin{array}{lrl}
{\rm Max =  HJD}& 2454236.7014 & + 0.12579931\,\, {\cdot\,\, E}\\
               &     \pm0.0001 &\pm0.00000011  
\end{array}
$\\
by means of a least--squares line.
The O-C values (differences between the observed and calculated times of maxima,
also listed in Table~2) were determined by using this ephemeris. 
The CCD $BVI$ measurements performed at the Piszk\'estet\H{o} Observatory in 2009 provide
an excellent validation. The maximum in $I$ light was observed at HJD~2455011.4997 when 
the above ephemeris predicts HJD~2455011.4994$\pm$0.0007 for E=6159.  The light curve
of this  maximum is 
flatter  than those of the $B$ and $V$ curves, whose maxima were observed at  HJD~2455011.4933 
and  HJD~2455011.4910, respectively.

\subsection{The independent terms}

Twelve independent frequencies were found in the interval 8.23--13.22~\cd, i.e.,
at periods shorter than that of  the fundamental radial mode \fu\, (Table~\ref{sol}). 
These additional modes could not be detected in the ASAS timeseries since the level of the
noise is much higher (0.020 and 0.025 mag in $V$ and $I$, respectively) than their amplitudes
(0.006 mag or less).

A close, but non--equidistant triplet of frequencies (\fd=10.48, \ft=10.69, and \fq=10.85~\cd, 
Fig.~\ref{ampli}, panel {\it b}) appears at an 
amplitude level similar to that of 4\fu, i.e., in the 3.0-6.2~mmag range. Is \hads\, a double--mode
HADS star? The frequency ratio between the
fundamental and first overtone modes in HADS stars is well constrained in the 0.771--0.774 interval
\citep[see Table~2 and Fig.~4 in][]{gsc}. In turn,
this implies that the first overtone of \hads\, is expected in the interval 10.27--10.31~\cd. None of the terms of
the triplet matches this value  and so the triplet is composed of nonradial modes.

The amplitude of \fq\, is half that of \ft, and  this ratio is also maintained for $f_5$, $f_6$
(Fig.~\ref{ampli}, panel {\it b}), and $f_7$. After $f_7$ (amplitude 0.27 mmag), the amplitude of the 
subsequent terms is slightly decreasing and we detected four other modes ($f_8, f_9, f_{10},$ and $f_{11}$) in
less than 0.10~mmag (Fig.~\ref{ampli}, panel {\it c}). Still, no regular separation
is observed among these  peaks. More generally, we did not find the same separation 
between two terms from \fu\, to $f_{13}$. Regarding the identification of the first radial overtone,
the $f_9$ term provides the ratio closest to the expected one, but it appears too discrepant
(see also Sect.~\ref{idmode}).

\subsection{The modulation of the \fu\, component}

The side peaks of \fu\, (7.7559 and 8.1424~\cd; Fig.~\ref{ampli}, panel {\it d}) and
2\fu\, (15.7051 and  16.0912~\cd)  
are the most surprising  features in the amplitude spectrum of \hads. They form an equidistant
triplet with a frequency separation \fB=0.193~\cd. In practice, this triplet is the signature
of a modulation of the main component \fu. The amplitudes of the modulation terms 
are much smaller than those of the independent terms discussed
above. This hampers any direct detection of the periodicity in the O--C or the
magnitude-at-maximum diagrams,
since the effects of the independent modes mask those of the modulation,
also taking the long (512~s) CoRoT exposures into account.  
The residual signal left around \fu\, (noted as ``Long term" 
in Fig.~\ref{ampli}, panel {\it d}) will be discussed in Sect.~\ref{fm1}.

The two frequencies $f_{12}$=9.325 and $f_{13}$=8.354~\cds are close, but not coincident,  
to the combination terms 
\fc--\fB=9.340~\cds and \fu+2\fB=8.335~\cd.  The hypothesis that they are actually 
combination terms implies an \fB\, value significantly
different from 0.193~\cds (0.208 and 0.202~\cd, respectively). Therefore, we should admit that
the  modulation is not the same  or that these two frequencies are independent
modes. Since the occurrence of a combination term such as \fc--\fB\, is quite unusual we prefer the
latter explanation.

\subsection{The combination terms}\label{duecd}
The amplitude spectrum of \hads\, is also rich in combination terms 
(Fig.~\ref{ampli}, panel {\it b}), as expected from the
amplitude of the independent terms from \fu\, to \fc, greater than 1~mmag.
The terms involving differences were also detected, and the lowest frequency is \fc$-$\fu=1.58~\cd.
We can argue that the bin on four pulsational cycles (i.e., 0.5~d) did not hamper
detection of peaks below 2~\cd, if any.
Patterns of combination terms with the harmonics up to 5\fu\ 
are detected between \fu\, and \fd, \fu\, and \ft, 
\fu\, and \fq, up to 2\fu\, with \fc\, and $f_6$. 
Only four frequencies (3\fu$-$\fd=13.00~\cd, 3\fu$-$\ft=13.37~\cd, 3\fu$-$\fq=13.16~\cd,
 and \fu+\fc=17.48~\cd) are within the interval 7-18~\cd.
We considered them as combination terms rather than independent modes, 
since they follow the patterns described above closely (see also Table~\ref{sol}).
We also detected the term \fd+\ft, i.e., the one involving the additional terms with the
largest amplitudes.
The modulation does not affect the 
combination terms; i.e., we did not detect any term having the form
\fu$\pm$\fB$\pm f_i$.
\section{The fit of the light curves in the different CoRoT passbands}
\begin{figure}[]
\begin{center}
\includegraphics[width=\columnwidth,height=0.5\columnwidth]{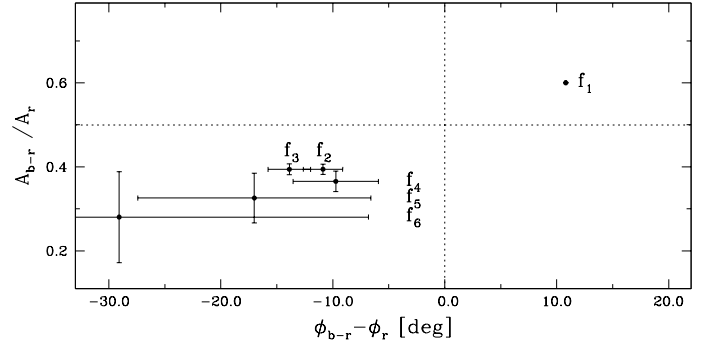}
\caption{\footnotesize Phase shifts and amplitude ratios of the frequencies from \fu\, to $f_6$\, as
 obtained from CoRoT coloured timeseries.
}
\label{grafo}
\end{center}
\end{figure}

Calculation of the phase differences and amplitude ratios in different
passbands could allow  the mode identification \citep{garrido}. 
Unfortunately, the CoRoT photometric system is not
a well--defined one, depending on the star itself \citep[see Sect.~7 in ][]{esa5}. 
To perform a comparative procedure, 
we calculated the least--squares fits of the coloured CoRoT and $BVI$ Konkoly timeseries.
The full amplitudes of the light variations of \hads\, are 0.432, 0.385, and 0.273~mag in
the $b_C$, $g_C$, and $r_C$  CoRoT passbands and   
0.459, 0.367, and 0.215~mag in the $BVI$ filters, respectively. Therefore,
the $b_C$ and  $g_C$ passbands are not separated well since both are between the $B$ and $V$ ones, while
the $b_C$ and $r_C$ passbands are both a little redshifted with respect to the $B$ and $V$ ones. 
We plotted   the phase shift $\phi_{b-r}-\phi_r$ vs.
the amplitude ratio $A_{b-r}/A_r$.  The position of the mode \fu\, is very
far from those of the modes \fd$-f_6$, all clustering in a  narrow region (Fig.~\ref{grafo}).  
Positive values of the phase differences suggest $\ell=0$ (i.e., radial) modes,
negative values are likely $\ell\geq1$ (i.e., nonradial) modes \citep{watson}. 
Moreover, we can see very good agreement between the observed 
positions of the modes \fd$-f_6$\, and 
the theoretical locus of the $\ell$=2 modes \citep[see Fig.~18 in ][]{watson}.
The identification of these frequencies with the $\ell$=2 modes must
be considered as only preliminary, due to the uncertainties 
in the characteristics of the CoRoT photometric system.

The phase errorbars associated to $f_6$\, are already too large to 
clearly point out the  mode identification, and this becomes impossible
for  the frequencies  from $f_7$\, to $f_{13}$.
In particular, we cannot obtain clues on the identification of 
$f_9$\, as the first radial overtone. 

\section{The stellar physical parameters}
 The  spectrum of 
\hads\,  was obtained  with the HARPS instrument mounted
at the 3.6-m ESO telescope on the nights of 29-30 June 2009.
Due to the star's faintness and the short periods,
the resulting SNR value was around 30 ($R$=80000, exposure time 2300~s),
and we could use the spectrum to estimate the physical parameters only.

The mean line profile was computed with the LSD software \citep{lsd}
and the Fourier transform of such profile supplied $V_{eq}\sin i$=26$\pm$1~\kms.
However, this determination is affected by the broadening owing to
the radial pulsation  of the \fu\, mode. 
Since we  do not know the amplitude of the radial velocity curve of \hads,
we estimated it by using the $2K/\Delta V$ value. For
radial pulsation it is around 100~\kms\,mag$^{-1}$ \citep{smith,v1719}, the highest
value was probably measured for V356 Aur \citep{aur}, i.e.,
130~\kms\,mag$^{-1}$. 
Since $\Delta V=0.37$ mag in the case of \hads, 
we got $2K$=37~\kms\, and 48~\kms, respectively. 
We also estimated  the $2K$ value in the
specific case of \hads, obtaining $2K$=12~\kms\, using the theoretical approach described by \citet{2K}.
The ephemeris determined in Sect.~\ref{ephem} allowed us to establish that
the HARPS spectrum 
started after the maximum light and ended around the halfway point of the 
descending branch.
In this interval the radial velocity due to
pulsation is getting close to zero after its maximum value.
We evaluated the effects of  both the radial pulsation and the exposure time 
on the mean line profile by using the software package FAMIAS \citep{famias}. 
Our simulations indicated that the $2K$ values determined above could produce a maximum broadening
of 5~\kms\, in the  determination of  the $V_{eq}\sin i$ value.
 By combining all these results, we accepted 
$23\pm4$~\kms\, as the value  of the $V_{eq}\sin i$ of \hads.

We computed a grid of synthetic spectra with the $R$=80000 resolution and the
observed line broadening by means of the ATLAS9 and SYNTHE codes \citep{synthe,atlas}.
The grid covers the ranges 7000$<T_{\rm eff}<$8000~K, 3.2$<\log g<$4.8, and --0.5$<$[Fe/H]$<$1.0. 
A microturbulent velocity of 2~\kms\, \citep{microt} was considered.
We compared the observed spectra with the synthetic grid in chosen
wavelength regions in order to estimate the physical parameters of the
star: the H$\alpha$ and H$\beta$ wings for $T_{\rm eff}$ and the Mg triplet
(5167, 5173, and 5184 \AA) for $\log g$.  We obtained finally
$T_{\rm eff}=7300^{-300}_{+200}$~K
and $\log g=3.75\pm0.20$ dex. These values were then used for a complete
abundance analysis \citep[see Sect.~4.2 in][]{50844}, 
which revealed a slight metallic solar underabundance, 
i.e.,  [Fe/H]=--0.15$\pm$0.20~dex \citep[{[Fe/H]}$_{\sun}$=7.54 dex;][]{asplund}.

Identifying the fundamental radial mode allows us to calculate the
absolute magnitude. We get $M_V$=1.46$\pm$0.10 by using the period--luminosity 
relation given by \citet {fornax}. Other relations
give similar results for $\log P=-0.900$ (see Fig.~8, ibidem).
The values  $M_V$=1.46  and $T_{\rm eff}= 7300$\,K puts \hads\,
in the middle of the HADS instability strip \citep[see Fig.~2 in][]{vienna}.

The approach described by \citet{bb} supplied the relation  
\begin{equation}
\log Q = -6.456 + \log P + 0.5 \log g + 0.1 M_{\rm bol} + \log T_{\rm eff}\,.
\end{equation}
We know that $P$=0.126~d from the CoRoT observations and that 
$Q$=0.033~d from the identification of $P$ as the fundamental radial mode. Since
the bolometric correction is $+0.03$~mag for $T_{\rm eff}=7300$\,K 
\citep{torres}, we obtain $M_{\rm bol}$=1.49$\pm$0.10 and $L=19.8~\pm1.8L_{\sun}$ 
from the  $M_V$=1.46 value calculated above.
The unknowns $g$ and $T_{\rm eff}$ must therefore satisfy the condition  
\begin{equation}
\log g + 2.0 \log T_{\rm eff}= 11.45(\pm0.01)\,. 
\end{equation}
This condition holds for the central values of $\log g$ and $T_{\rm eff}$ obtained from HARPS
spectra and proves the self--consistency of the physical model. We can then calculate the
radius $R=2.8\pm0.4~R_{\sun}$ and the mass $M=1.6\pm0.7~M_{\sun}$ from the $L$, $T_{\rm eff}$, 
and $\log g$ values.

We further checked the consistency of the physical model by combining our  
$BV$ photometry with  the 2MASS photometry.
The mean $BV$ magnitudes  are $<V>$=13.37 and $<B>$=13.94. These
values are in good agreement with those provided by the EXODAT catalogue \citep[$V$=13.43, 
$B-V$=+0.60;][]{exodat}. Indeed, the epoch of the EXODAT observations corresponds to the
phase 0.23~p, i.e., roughly the middle of the descending branch (Fig.~\ref{f1}). We calculated
the equivalent widths (EWs) of the NaI lines at 5890.0 and 5895.9\,\AA\, from the HARPS spectrum.
We could estimate the
interstellar reddening $E_{B-V}$=+0.20 by means of the $E_{B-V}-$EW relationship \citep{sodio}, also 
noticing that the interstellar lines are characterized by multiple components,  one of which is
in emission.
In turn,  the $E_{V-K}$=+0.56 value was calculated using the relation given by \citet{vk}, hence
$(V-K)_0$=+0.70 from the 2MASS value $K$=12.11 \citep{2mass}.
The latter index  supplies a $T_{\rm eff}$ value of 7250~K 
\citep[Eq.~8 in][]{masana}, in perfect agreement with the one  provided by spectroscopy.
Table~\ref{pp} lists the physical parameters obtained from space--based photometry
and ground--based spectroscopy.

\begin{table}
\setcounter{table}{2}
\caption{Observed and physical parameters of \hads\, determined
in this paper.}
\begin{tabular} {lc c| lc}
\hline
\hline
\multicolumn{1}{c}{Parameter} &  \multicolumn{1}{c}{Value} && \multicolumn{1}{c}{Parameter} & \multicolumn{1}{c}{Value}\\
\hline
\noalign{\smallskip}
$<V>$                &      13.37       &&    $T_{\rm eff}$ $^{\mathrm{b}}$     &   7300$^{-300}_{+200}$ K \\
$<B>$                &      13.94       &&    $\log g $          &    3.75$\pm$0.20  \\
$E_{B-V}$            &      +0.20       &&    {[Fe/H]} $^{\mathrm{c}}$   &   --0.15$\pm$0.20 dex  \\
$K^{\mathrm{a}}$ &      12.11       &&    $V_{eq}\,\sin i$   &    23$\pm$4 \kms \\   
$(V-K)_0$            &      +0.70       &&    $L$                &    19.8$\pm$1.8 $L_{\sun}$  \\
$P_{\rm fund}$     &    0.1258 d        &&    $R$                &    2.8$\pm$0.4 $R_{\sun}$   \\
$M_V$              &    1.46$\pm$0.10   &&    $M$$^{\mathrm{d}}$ &    1.71$\pm$0.02 $M_{\sun}$   \\
$M_{\rm bol}$      &    1.49$\pm$0.10   &&    Age$^{\mathrm{d}}$ &    1485$\pm$10 Myr      \\
\hline
\end{tabular}
\label{pp}
\begin{list}{}{}
\item[$^{\mathrm{a}}$] From \citet{2mass}
\item[$^{\mathrm{b}}$] Pulsational models suggest $T_{\rm eff}$=7000$\pm$50 K
\item[$^{\mathrm{c}}$] Pulsational models suggest [Fe/H]=--0.30$\pm$0.20 dex
\item[$^{\mathrm{d}}$] Values from pulsational models
\end{list}
\end{table}

\begin{table}
  \begin{center}
   \caption{Mode identification from  the best pulsational model.}  
   \begin{tabular}[]{l rr rlr r}
   \hline
   \hline
\noalign{\smallskip}
Term  &  \multicolumn{2}{c}{Obs. freq.} & \multicolumn{3}{c}{Mode ident.} & Calc. freq. \\
      &  \multicolumn{1}{c}{[\cd]}&\multicolumn{1}{c}{[$\mu$Hz]}  & \multicolumn{1}{c}{$n$} & \multicolumn{1}{c}{$\ell$}  &
\multicolumn{1}{c}{$m$} & \multicolumn{1}{c}{[$\mu$Hz]} \\
\noalign{\smallskip}
   \hline
\noalign{\smallskip}
$f_{1}$  &  7.9492 &  92.004 &     $     1 $ &   0 & $   0  $  &     92.164  \\
$f_{2}$  & 10.8470 & 125.544 &     $    -2 $ &   2 & $   0  $  &    126.228  \\
$f_{3}$  & 10.4795 & 121.291 &     $     1 $ &   1 & $   0  $  &    121.489  \\
$f_{4}$  & 10.6903 & 123.730 &     $     1 $ &   1 & $  -1  $  &    123.114  \\
$f_{5}$  &  9.5326 & 110.331 &     $    -8 $ &   5 & $   2  $  &    110.432  \\
$f_{6}$  & 13.1759 & 152.498 &     $     2 $ &   1 & $   0  $  &    153.024  \\
$f_{7}$  &  9.7961 & 113.381 &     $    -3 $ &   2 & $   1  $  &    113.369  \\
$f_{8}$  &  8.2326 &  95.284 &     $    -1 $ &   1 & $   1  $  &     95.047  \\
$f_{9}$  & 10.2033 & 118.094 &     $    -3 $ &   2 & $  -2  $  &    118.367  \\
$f_{10}$ & 12.6351 & 146.240 &     $    -2 $ &   3 & $   2  $  &    145.999  \\
$f_{11}$ & 13.2246 & 153.063 &     $    -1 $ &   3 & $   2  $  &    153.356  \\
$f_{12}$ &  9.3252 & 107.931 &     $    -10 $ &  5 & $   1  $  &    108.123  \\
$f_{13}$ &  8.3543 &  96.693 &     $    -2 $ &   1 & $   0  $  &     96.832  \\
\noalign{\smallskip}
    \hline        
\end{tabular}
   \label{modes}
  \end{center}
\end{table}

\begin{figure}[t]
\begin{center}
\includegraphics[width=\columnwidth,height=\columnwidth]{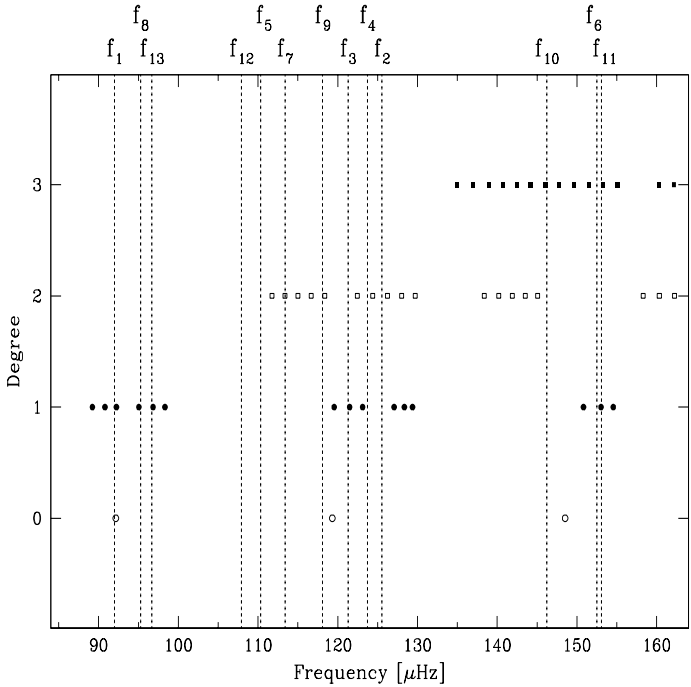}
\caption{\footnotesize 
{The matching between observed (vertical lines) and calculated (points) frequencies for models
with degree $\ell\le~3$.}}
\label{elle}
\end{center}
\end{figure}

\section{Pulsational model and  mode identification} \label{idmode}

The parameters obtained from CoRoT photometry and HARPS spectroscopy
were used to calculate a preliminary  pulsational model, 
making certain to match the constraint that \fu\, must be a radial mode. 
We  followed 
the procedure and the physical assumptions described in \citet{casas}.
In particular, we considered a variation in the
convective efficiency parameter $\alpha_{MLT}$ up to 1.5, of the
overshoot $d_{ov}$ up to 0.3, and  an [Fe/H] metallicity in the range from --0.52 to
0.08 dex. Other effects such as diffusion and radiative forces were not included.
We assumed a uniform rotation since we could not identify multiplets in advance
among the 13 frequencies.
A non-uniform rotation 
would be  an additional source of uncertainty since
mixed modes are highly sensitive to the rotation profile 
near the stellar core \citep{pet2,nueri}.

This procedure points out that a further refinement could be performed 
starting from 
$M=1.71\pm0.02~M_{\sun}$, $L=18.5\pm0.2~L_{\sun}$, $R=2.90\pm0.04~R_{\sun}$,
$T_{\rm eff}=7000\pm50$~K, $\log g=3.74\pm0.01$~dex, and [Fe/H]=--0.30$\pm$0.20 dex.
These results indicate  a cooler $T_{\rm eff}$ and  more pronounced metallic underabundance
than those suggested by the spectroscopic analysis, though
the errorbars still overlap in both cases. Models with $T_{\rm eff}\ge 7300$~K 
are not able to fit \fu\, as the fundamental radial mode and they moreover identify
the other frequencies as gravity modes. In a similar way, solar abundances 
lead to massive, main-sequence stars. All these possibilities 
are unlikely when considering the whole set of parameters and the 
observed photometric variability. 

We should take into account that the 
$V_{eq}\sin i$ value suggests a non-negligible rotational velocity.
The adiabatic oscillation code  FILOU 
\citep[][and references therein]{phd,Sua08filou} computes the oscillation
frequencies up to the second-order effects of rotation. 
It includes the corrections of the frequencies near degeneracy effects
occurring when two frequencies are close to each other.
These effects could also occur between
three or more modes, but they only should be relevant for very high
rotational velocities  \citep{Sua06rotcel,Sua10solar}.

Starting from the reference model pointed out by the spectroscopic and photometric data,
we computed sets of models with different rotational velocities, ages,
and maximum degree $\ell$ of the modes by using the code FILOU.  
We had to make some assumptions on the inclination angle $i$ in order to  
constrain the rotational frequency and then infer the rotational splitting.  
The $V_{eq}\sin i$ value  of HADS stars is usually below 25~\kms\,
\citep[Fig.~5 in][]{brevie},  suggesting that these variables are intrinsically
slow rotators. Thus, 
we considered \hads\, as seen nearly  equator-on, and we  allowed  variations
of $V_{eq}$ in the range 19--30~\kms. 

The oscillation spectra predicted by the models 
were compared with the 13 independent terms detected in the frequency analysis. 
The pulsational models limited to $\ell\le~3$  directly  match 11 out of the 13 frequencies.
 Indeed, modes are predicted around all
the observed frequencies, except for \fc\, and $f_{12}$ (Fig.~\ref{elle}). To fill this gap,
we computed  the models  with $\ell\le~5$ and could obtain 
the  mode identifications for these frequencies, too (Table~\ref{modes}).
The average difference between observed and calculated frequencies 
is $0.28\,\mu$Hz.
The model  depends on our assumptions and 
on the uncertainties in the identifications of close frequencies; 
therefore they cannot be considered unique.  We note that
identifying all frequencies at once with the modes predicted by the models
with $\ell\le~5$ improved the fit, because the oscillation spectrum would be denser. 
However, the high prevalence of  $\ell\ge~4$ modes on $\ell\le~3$ ones 
is contrary to what is expected, since  
the amplitudes of the modes should decrease with increasing $\ell$
due to the cancellation effects. 

The mode identification describes a star  very close to  the terminal age on the main sequence,
with a central hydrogen content $X_{\rm c}=0.15\pm0.05$ and, in turn, an age of $1485\pm10$~Myr. 
This evolutionary stage agrees with the 
identification of the additional terms (i.e., all the terms except \fu)
with mixed modes. The $f_9$ mode is not identified with the first radial overtone,
confirming that the \fu$/f_9=0.779$ value differs too much from the predicted value 
(0.774).

We tried a fine tuning of the model after noticing 
that the  \ft\, and \fq\, terms are identified as two components of a rotationally
split $\ell=1$ multiplet (Table~\ref{modes}). We considered  the existence of such  multiplet
as a further input constraint.
To do this, we computed models with small variations in the rotational velocity
around the observed $V_{eq}\sin i$. Such variations had to be large enough to
modify the  $m\neq0$ components sufficiently, and small enough to avoid
variations in the $m=0$ component.
The latter constraint allows a variation of only about 4~\kms.
Within this  range, the second-order effects of rotation do not change the identification of 
the other modes. Therefore, the model suggests that the 
\fd\, mode definitely does not belong to the 
same multiplet as  the \ft\, and \fq\, modes.

\begin{figure}[]
\begin{center}
\includegraphics[width=\columnwidth,height=0.5\columnwidth]{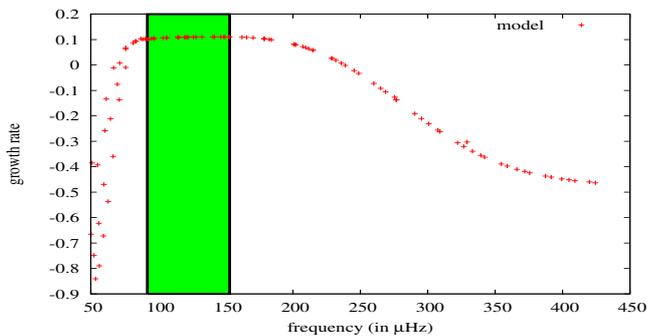}
\caption{\footnotesize 
Energy balance of the modes of the best fitting model. 
The green box represents where the observed modes are located in frequency. 
}
\label{andy}
\end{center}
\end{figure}

As a last step,  
the non-adiabatic analysis of the mode energy balance 
was performed by means of the  GraCo code \citep{moya1,moya2}. Figure~\ref{andy}
shows the growth range of the model. Overstable modes must have a positive growth rate,
and this happens for a region including the observed frequencies.

Improvements in the theoretical treatment of the mixed modes can lead to  a 
more accurate fitting of the oscillation spectrum of \hads.
In any case the mixed modes
can provide significant constraints on the stellar internal structure since now.
In particular, our  model predicts 
a convective core with  size $r_{\rm c}\sim0.08\,R_{\sun}$ and a
thin outer convective zone located at $r = 0.99\,R_{\sun}$, 
as well as $\alpha_{MLT}$=0.5,
in agreement with \citet{casas}, and  $d_{ov}=0.1$.
These two values affect not only the
size of the convective zones, but also the possibility of estimating the
time spent  on the main sequence.

\section{Discussion}

The CoRoT satellite offered the possibility of monitoring  a
variable star belonging to the class of HADS stars in a continuous and long way. At the
moment, the pulsational behaviour disclosed by the timeseries of \hads\, is the most 
advanced observational fact we have to understand 
the physical mechanisms at work in the variables of this class.
These are some questions raised by previous ground--based
observations on which the CoRoT photometry shed a new light.

\subsection{The excitation of independent modes}
The excitation of small--amplitude nonradial modes accompanying the high--amplitude radial
modes in  $\delta$ Sct variables has been debated at length. The pulsational content
of V974 Oph provided the decisive  clear piece of evidence, since  this star
pulsates in the fundamental radial mode ($f_1$=5.23~\cd, full amplitude 0.32 mag in $B$)
and in at least four other independent modes easily  detectable also from
ground--based timeseries \citep[e.g., $f_2$=5.35~\cd, full amplitude 0.14 mag  in $B$--light;][]{oph}. 
The frequency ratios $f_1/f_i$ of these  additional modes 
(0.734, 0.786, 0.807, and 0.978) clearly indicate that most of them, if not all,
are nonradial modes.
In the case of \hads\, the amplitudes of the nonradial modes are much smaller than those
of V974 Oph, but the superior quality of the CoRoT timeseries does not leave any ambiguity
in their  identification. The ratios \fu/$f_i$ are listed in Table~\ref{sol}. The ratios
0.733, 0.811, and 0.779 are very similar to the 0.734, 0.807, and 0.786 ones obtained in
the case of V974 Oph. However, the corresponding modes have a different rank in amplitude in
the two solutions of the light curves of the two stars. 

We can also compare the pulsational  content  of \hads\, with that of 
HD~50844. In this low--amplitude $\delta$ Sct star, we detected an increase in 
the signal at $f<30$~\cds and hundreds
of frequencies were necessary to reduce it to  the same level of the noise observed
at $f>30$~\cds \citep{50844}. After removing the peaks related to the 61 terms
detected in the CoRoT data of \hads,  the noise level is constant over the full 
5-100~\cds spectrum (see inserted box in panel {\it e}  of Fig.~\ref{ampli}).
This noise level (29~$\mu$ mag) after subtracting 61 frequencies is much 
less  than observed in the 
amplitude spectrum of HD~50844 after subtracting 500 frequencies \citep[see Fig.~3 in][]{50844}.
Moreover, the amplitude of 0.10~mmag is reached after detecting of about 250
frequencies in the light curve of HD~50844, against the only 61 ones in the case 
of \hads.
We can conclude that the variability of \hads\, is much simpler than that of
HD~50844. New cases will allow us to verify if the excitation of only one ten of  
modes is a characteristic of the HADS stars.

\subsection{The  modulation of \fu}\label{fm1}

The physical interpretation of the \fB\, term is not an easy task.
We could suppose that it is related with the star's rotation, so $P$=5.181~d
is the rotational period. Under this assumption the radius
value listed in Table~\ref{pp} supplies an equatorial velocity $V_{eq}=27\pm4~$\kms,
supporting the hypothesis that  \hads\, is seen nearly equator-on. 
This scenario seems coherent, but we have to explain why
the rotation affects the amplitude of the main pulsation term. 
A chemical inhomogeneity on the surface could do it, but then it should be detected
as an independent peak in the amplitude spectrum. Moreover, the \fB\,
harmonics should also be detected, since
the equator-on orientation  makes the continuous visibility unlikely. 
Actually, frequency \fB\, is not observed in the amplitude spectrum of \hads.
This lack could be due to both its intrinsic, very small amplitude and
the instrumental effects that complicate the spectrum at $f<1$~\cd.

A continuous modulation in amplitude and/or in phase superposed the main 
pulsational period is commonly observed in RR Lyr stars, and it defines
the Blazhko effect. 
The onset of the Blazhko modulation is due to something 
more specific than rotation, e.g.,  
the resonance model between nonradial, low--degree modes and the main radial
mode \citep{dzie}, or the oblique pulsator  model in which the rotational
axis does not coincide with the magnetic axis  \citep{shiba}
or the action of a turbulent convective dynamo in the lower envelope of the star \citep{sto}.
\citet{blabre} discusses the possibility of explaining the amplitude and phase modulations
observed in three low--amplitude $\delta$ Sct stars (FG Vir, BI CMi, and AI CVn)
as due to the Blazhko effect. The very rich pulsational content of these pulsators
favours the interaction between modes, without invoking the 
resonance mechanism suggested by \citet{dzie}. In the case of HADS stars, a much
smaller number of modes are excited, and then it appears more obvious to link \hads\, with
RR Lyr stars than with low--amplitude $\delta$ Sct stars, also
having the sharp triplet structure in mind.

It could be useful to investigate the light curves under the magnification
lens provided by  the CoRoT timeseries.
The long--term stability of the Blazhko modulation in RR Lyr stars is a new issue opened up
by the continuous CoRoT timeseries \citep{793}. 
In the case of the HADS variable \hads, there are  very low peaks 
at \fu\, and harmonics after subtracting the 61 terms
(Fig.~\ref{ampli}, panel {\it d}), but the instrumental
effects revealed by the analysis of the chromatic data can be responsible for this
residual signal. The amplitudes of \fu\, in white light calculated
in each bin are slightly varying during the CoRoT monitoring
(bottom panel in Fig.~\ref{f1}), but this effect is very marginal and could still reflect
the complicated  changes in the trends of the $b_C$, $g_C$, and $r_C$ data.
The light curves of Blazhko RR Lyr stars show bumps and larger scatter 
at well--defined phases  \citep[e.g., Fig.~3 in][]{793}, probably in
relation with shock waves.  In the case of \hads,
the smoothed  light curve over \fu\, (Fig.~\ref{f1}, upper curve in the top panel)
and the residuals folded with \fu\, (Fig.~\ref{f1}, lower curve in the top panel)
do not show any particularity.
Therefore, at the moment we have no further evidence
supporting the straight identification of the triplets 
around \fu\, as due to the  Blazhko effect observed in RR Lyr stars. 

In RR Lyr stars, the amplitude of the Blazhko modulation is stronger than that
of additional modes. Indeed, CoRoT photometry allowed us to study the Blazhko
effect in great detail  and the discovery of  additional modes was a bonus \citep{aql,793}. 
In the case of HADS stars, the opposite case occurred; i.e., searching for additional modes 
we discovered  the \fu\, modulation. This demonstrates once more how rewarding 
the investment in the observation  of known phenomena with more advanced tools could be.
  
\section{Conclusions}
The solution for the CoRoT light curve of \hads\, allowed us to detect only 12 additional
modes;
i.e., we did not detect the huge number of frequencies found in low--amplitude 
pulsators \citep{50844}.  These modes all have mixed nature and, together with
the fundamental radial mode, originate sixty-one frequencies able to solve  the
CoRoT light curve. We can speculate that 
the large amount of energy put in the radial modes limits the number of excited 
nonradial modes, but the observations of new HADS stars could shed new light on
this particular aspect.

The newest result is the small but very clear modulation of the fundamental radial mode,
observed for the first time in an HADS star. The observation of other HADS stars by
space photometry could tell us whether this is a rotational effect
or an extension of the Blazhko effect from RR Lyr to HADS stars. 
The similarity between these two classes of variables would be  
a new way of looking at the Hertzsprung-Russell diagram and, if confirmed, 
would be able to put 
tighter constraints on the physical model of the Blazhko effect, also taking
the thin convective layer predicted by the model of \hads\, into account. Indeed, it should be
able to explain the modulations observed both in low--mass stars in a core He--burning phase
(RR Lyr stars) and in more massive, post-main sequence stars moving to the giant
branch (HADS stars).  In this context, nonradial resonant modes close to the radial
fundamental mode seem to have the requisites needed to act in such different
evolutionary stages, but   the other models are still plausible.
We also note that the modulation observed in \hads\, (5.181~d) is shorter in time
than the shortest ever 
observed for Blazhko RR~Lyr stars \citep[5.309~d in SS~Cnc;][]{sscnc}. 

The excitation of mixed modes in stars approaching the terminal age on the main sequence
is confirmed and they are found in the predicted overstable region.  
Further improvements in the theoretical modelling are necessary, especially
when dealing with close frequencies in the presence of a non--negligible rotation.
\begin{acknowledgements}
We acknowledge the support from the {\it Centre National
d'Etudes Spatiales} (CNES).
The spectroscopic  data were obtained as part of the ESO Large Programme
LP182.D-0356 (PI.: E.~Poretti) in the framework of
the Italian ESS project, contract ASI/INAF~I/015/07/0, WP 03170.
The authors wish to thank the anonymous
referee for useful comments.
EN acknowledges financial support of the NN203 302635 grant from the
MNiSW.
MP, JMB, and ZsB acknowledge the support of the ESA PECS projects No.
98022 \& 98114.
AM acknowledges the funding of AstroMadrid (CAM S2009/ESP-1496).
JCS acknowledges support from the Instituto de Astrof\'{\i}sica de
Andaluc\'{\i}a (CSIC) by an Excellence Project postdoctoral fellowship
financed by the Spanish {\it Conjerer\'{\i}a de Innovaci\'on, Ciencia y Empresa
de la Junta de Andaluc\'{\i}a} under project FQM4156-2008. JCS also
acknowledges support by the Spanish {\it Plan Nacional del Espacio} under
project ESP2007-65480-C02-01.
We thank 
J.~Vialle for checking the English form of the original manuscript.
This publication makes use of data products from the Two Micron All Sky Survey, 
which is a joint project of the University of Massachusetts and the Infrared 
Processing and Analysis Center/California Institute of Technology, 
funded by the National Aeronautics and Space Administration and the National Science Foundation.

\end{acknowledgements}

\end{document}